\documentclass[journal, 10pt]{IEEEtran}
\IEEEoverridecommandlockouts

\makeatletter
\let\NAT@parse\undefined
\makeatother

\usepackage{amsmath,amssymb,amsfonts,amsthm,commath,esint}
\usepackage{bm} 
\usepackage[numbers,sort&compress]{natbib}

\usepackage{mathtools}
\usepackage{multirow}
\usepackage{physics}
\usepackage{color} 
	
\usepackage{colortbl}
\definecolor{Gray}{gray}{0.9}

\usepackage{graphicx}  
\usepackage{enumitem}  
\usepackage{optidef} 
\usepackage{mleftright}
\usepackage{cuted}
\usepackage{algorithm2e}
\usepackage{microtype}
\usepackage{ifthen} 
\usepackage{subcaption}
\usepackage{soul}
\usepackage{array} 

\SetKwInput{kwInit}{Initialization}
\ifCLASSOPTIONcompsoc
    \usepackage[caption=false, font=normalsize, labelfont=sf, textfont=sf]{subfig}
\else
\usepackage[caption=false, font=footnotesize]{subfig}
\fi

\makeatletter
\def\blfootnote{\xdef\@thefnmark{}\@footnotetext}
\makeatother

\newcommand{\p}[1]{\left(#1\right)}

\renewcommand{\vec}[1]{\mathbf{\lowercase{#1}}}

\newcommand{\mat}[1]{\mathbf{\uppercase{#1}}}

\newcommand{\veci}[1]{\pmb{\lowercase{#1}}}

\newcommand{\T}{^\mathsf{T}}    
\renewcommand{\H}{^\mathsf{H}}   

\newcommand{\e}[1]{\mathrm{e}^{#1}}

\DeclarePairedDelimiterXPP\Aver[1]{\mathbb{E}}{[}{]}{}{

#1
}
 
\definecolor{purple}{rgb}{0.63, 0.13, 0.49}
\definecolor{sepia}{rgb}{0.44, 0.26, 0.08}
\definecolor{pink}{rgb}{1, 0.41, 0.79}
\definecolor{dkgreen}{rgb}{0.4, 0.7, 0.2}
\definecolor{blue}{rgb}{0.0, 0.0, 1.0}

\begin{document}

\title{{Digital Twin-Assisted Measurement Design and Channel Statistics Prediction}

\thanks{ 
        \textsuperscript{\text{*}}Robin J. Williams and Mahmoud Abouamer contributed equally to this work. This work was supported by the Villum Investigator Grant ``WATER" from the Velux Foundation, Denmark.
    }}
\author{Robin J. Williams\textsuperscript{\text{*}},~Mahmoud Saad Abouamer\textsuperscript{\text{*}},~and~Petar Popovski~\\
    Department of Electronic Systems, Aalborg University, Denmark\\
    Email: $\{$rjw,~mahmoudabo,~petarp$\}$@es.aau.dk}

\maketitle
\thispagestyle{empty} 

\begin{abstract}
Prediction of wireless channels and their statistics is a fundamental procedure for ensuring performance guarantees in wireless systems. Statistical radio maps powered by Gaussian processes (GPs) offer flexible, non-parametric frameworks, but their performance depends critically on the choice of mean and covariance functions. These are typically learned from dense measurements without exploiting environmental geometry. Digital twins (DTs) of wireless environments leverage computational power to incorporate geometric information; however, they require costly calibration to accurately capture material and propagation characteristics. This work introduces a hybrid channel prediction framework that leverages uncalibrated DTs derived from open-source maps to extract geometry-induced prior information for GP prediction. These structural priors are fused with a small number of channel measurements, enabling data-efficient prediction of channel statistics across the entire environment. By exploiting the uncertainty quantification inherent to GPs, the framework supports principled measurement selection by identifying informative probing locations under resource constraints. Through this integration of imperfect DTs with statistical learning, the proposed method reduces measurement overhead, improves prediction accuracy, and establishes a practical approach for resource-efficient wireless channel prediction.
\end{abstract}



\section{Introduction}\label{sec:introduction}

Accurate characterization of spatially varying channel statistics is essential for ultra-reliable low-latency communication (URLLC). In particular, stringent latency constraints limit the feasibility of extensive channel state information (CSI) acquisition and reduce the practicality of feedback-heavy mechanisms. In such settings, prediction based on spatial channel statistics becomes an attractive alternative, enabling robust rate adaptation that meets reliability constraints with high probability. This motivates the use of statistical radio maps~\cite{GP_radio_map_2}.

State-of-the-art approaches to radio map construction rely on non-parametric models, with Gaussian processes (GPs) providing a powerful framework for channel prediction. By modeling spatial correlations through mean and covariance functions, GPs enable interpolation of channel statistics at unmeasured locations without requiring explicit fading assumptions such as Rayleigh or Rician models \cite{GP_radio_map_3}. The effectiveness of GPs for statistical channel prediction has been validated in both simulations and real-world deployments \cite{GP_radio_map_3, GP_radio_map_2}. A key challenge for existing GP-based methods, however, lies in their reliance on a large number of channel measurements to learn accurate prior models. Since geometric structure is typically ignored, mean and covariance functions must be inferred directly from sampled channel statistics, leading to a dense offline data collection burden.  

To address these limitations, statistical radio maps can be augmented with digital twins (DTs) of wireless environments. A DT constructed from environmental geometry offers the potential to develop site-specific channel models that enable accurate prediction \cite{survey_DT}. When carefully calibrated by tuning material parameters to align simulated and empirical channels, DTs can serve as surrogates for direct channel prediction. Recent advances in differentiable ray tracing, such as SionnaRT \cite{Sionna_Intro}, have further enabled data-driven calibration through gradient-based learning of material properties. A growing body of work on DT calibration \cite{Calibration_Advance} and robust optimization using DT information \cite{DT_robust_optimization} highlights the promise of this approach. However, accurate calibration remains computationally expensive and difficult to maintain in dynamic environments where propagation conditions evolve over time. This challenge is particularly pronounced for URLLC applications, where tight latency constraints make direct prediction based on calibrated DTs impractical. 

To address these challenges, our earlier work \cite{GP_radio_map_5} incorporated uncalibrated DT-derived features into GP-based radio maps, demonstrating improved channel statistics prediction compared to conventional GP models. However, this feature-based integration treats the DT as an auxiliary enhancement and maintains a clear separation between the DT and the probabilistic inference model. As a result, DT-induced uncertainty is not fully exploited to improve prediction accuracy or to guide the selection of spatial locations at which channel measurements are collected.

In this work, we propose a hybrid channel prediction framework that combines the expressive power of GPs with geometry-induced priors obtained from uncalibrated DTs.  Consequently, the framework does not rely on calibrated DTs or accurate material knowledge. Instead, DTs constructed directly from satellite images or open-source maps such as OpenStreetMap \cite{OpenStreetMaps} are used to capture the dominant geometric structure of the environment. These geometry-induced priors are fused with a small number of channel measurements, enabling data-efficient spatial prediction of channel statistics. Conceptually departing from \cite{GP_radio_map_5}, this work treats the uncalibrated DT not merely as a feature generator, but as a probabilistic source of geometry-aware prior information for both the GP mean and the covariance functions. Furthermore, the proposed DT-informed GP framework enables informed selection of the spatial locations at which channel measurements are collected and subsequently used for GP-based prediction. Specifically, this work:
\begin{itemize}
    \item leverages uncalibrated DTs to construct geometry-aware GP priors that capture environment-induced correlations and enable reliable prediction across the entire region from a limited number of channel measurements
    \item designs optimized measurement strategies that exploit GP-based uncertainty quantification to select the most informative measurement locations, minimizing prediction error under strict measurement budget constraints
\end{itemize}
Numerical evaluations demonstrate that the proposed framework consistently outperforms geometry-unaware benchmarks as well as DT-assisted approaches that do not explicitly integrate DT-derived information into the probabilistic inference process, achieving lower channel statistics prediction error and improved statistical guarantees for URLLC rate selection.

\section{System Model and Scenario} \label{sec:system_model}

We consider a wireless communication scenario where an access point (AP) located at position $\vec{x}_{\text{AP}} \in \mathbb{R}^3$ serves users in a service region $\mathcal{R} \subseteq \mathbb{R}^3$.  
A single-antenna user transmits a zero-mean, unit-power symbol $s \in \mathbb{C}$ with transmit power $P_{\text{tx}} \geq 0$.  
The received signal at the AP is
\begin{align}
y = \sqrt{P_{\text{tx}}}\, h(\vec{x})\, s + n ,
\end{align}
where $h(\vec{x}) \in \mathbb{C}$ is the complex channel coefficient at location $\vec{x}$, and $n \sim \mathcal{CN}(0,\sigma_n^2)$ is additive white Gaussian noise.  
The  SNR is $\gamma(\vec{x}) = \frac{P_{\text{tx}} |h(\vec{x})|^2}{\sigma_n^2},$
and the channel power is $p(\vec{x}) = |h(\vec{x})|^2$.  The $\epsilon$-quantile of $p(\vec{x})$,
\begin{align}
p_\epsilon(\vec{x}) \triangleq \sup \left\{ t \in \mathbb{R}_+ \,\big|\, \Pr\!\left( p(\vec{x}) \leq t \right) \leq \epsilon \right\},
\end{align}
is of particular interest as it captures rare but critical channel behavior that may cause outages, and is therefore essential for ultra-reliable low-latency communications (URLLC).\footnote{{Although we motivate our framework using the channel power $p(\vec{x})$, the proposed methods apply more generally to other location-dependent channel metrics. Moreover, while $p(\vec{x})$ is introduced via a single-user signal model, it is defined over space and can be evaluated at multiple locations, enabling the estimation of statistics for arbitrary numbers of users.
}
}  

For instance, $p_\epsilon(\vec{x})$ can be used for URLLC rate selection by choosing the maximum rate $R(\vec{x})$ such that the outage probability is at most $\epsilon \in (0,1)$.  
Defining the outage probability as $p_{\text{out}, \vec{x}} \overset{\Delta}{=} \Pr \big( \log_2(1 + \gamma(\vec{x})) < R(\vec{x}) \big),$
the rate $R(\vec{x})$ must satisfy the meta-probability constraint
\begin{align}
\label{ref: meta prob 1}
\tilde{p}_{\epsilon} \triangleq \Pr \left( p_{\text{out}, \vec{x}} > \epsilon \right) \leq \delta .
\end{align}
Equivalently, one can express this condition as \cite{GP_radio_map_2}
\begin{align}
\label{ref: meta prob 2}
 \tilde{p} _{\epsilon} = \Pr \left( R(\vec{x}) > \log_2 \!\Big(1 +  \tfrac{P_\text{tx} \, p_\epsilon(\vec{x})}{\sigma_n^2} \Big) \right) \leq \delta,   
\end{align}
and hence rate selection for URLLC depends on the knowledge of $p_\epsilon(\vec{x})$. 
A direct way to estimate the $\epsilon$-quantile $p_\epsilon(\vec{x})$ at location $\vec{x}$ is to collect many samples of $p(\vec{x})$.  
However, for small $\epsilon$, the required sample size can become prohibitive due to latency or user mobility.  
To overcome this, we exploit spatial correlation in the wireless channel statistics (i.e. $p_\epsilon(\vec{x})$), over a region of interest $\mathcal{R}$.  

\section{Problem Statement and Design Objectives}
\label{sec:Problem Statement}
We model $p_\epsilon(\vec{x})$ as a spatial process and aim to estimate  
\begin{align}
\label{transform_stat}
t(\vec{x}) \triangleq g\!\left(p_\epsilon(\vec{x})\right),
\end{align} 
from a small number of measurements. 
Here the statistic of interest is defined as a transformation of $p_\epsilon(\vec{x})$, and in this work we choose $g(\cdot)=\log(\cdot)$. 
The motivation for this choice is discussed in Section~\ref{subsection:processing}.
Towards this, we employ GPs to model 
\begin{align}
  t(\vec{x}) \sim \mathcal{GP} \big( m(\vec{x}), c(\vec{x}, \vec{x}') \big),  
\end{align}
where $m(\vec{x}) = \mathbb{E}[t(\vec{x})]$ is the mean function and 
$c(\vec{x}, \vec{x}') = \mathbb{E}[(t(\vec{x}) - m(\vec{x}))(t(\vec{x}') - m(\vec{x}'))]$ is the covariance function.  
The expressive power of the GP framework depends critically on the appropriate specification of $m(\vec{x})$ and $c(\vec{x}, \vec{x}')$.

\subsection{From Gaussian Processes to Finite-Dimensional Inference}  

An important property of GPs is that, although they define distributions over an infinite number of random variables, any finite collection of these variables follows a multivariate Gaussian distribution.  
Formally, for any finite subset $\mathcal{V} \subset \mathcal{R}$, the random vector $\mathbf{t}_{\mathcal{V}} \triangleq \{ t(\vec{x}) : \vec{x} \in \mathcal{V} \}$
is distributed as 
\begin{align}
 \mathbf{t}_{\mathcal{V}} \sim \mathcal{N}\!\big( \mathbf{m}_{\mathcal{V}}, \, \mathbf{C}_{\mathcal{V}\mathcal{V}} \big),   
\end{align}
where $\mathbf{m}_{\mathcal{V}}$ is the mean vector with entries $m(\vec{x})$ for $\vec{x} \in \mathcal{V}$, and $\mathbf{C}_{\mathcal{V}\mathcal{V}}$ is the covariance matrix with entries $c(\vec{x}, \vec{x}')$ for $\vec{x}, \vec{x}' \in \mathcal{V}$.  

This connection allows us to leverage the infinite-dimensional GP model in practical inference tasks by restricting attention to a finite set of candidate prediction locations.  
In particular, we define a finite grid of positions $\mathcal{S} \subset \mathcal{R}$, which we refer to as the \emph{prediction space}.  
Our ultimate goal is to predict the value $t(\vec{x}^*)$ for some $\vec{x}^* \in \mathcal{S}$, using both the prior $(m(\cdot), c(\cdot,\cdot))$ and a set of noisy measurements collected at a subset of locations.  

\paragraph{Measurements at a subset of locations}  
Let $\mathcal{A} \subseteq \mathcal{S}$ denote the set of measurement locations, and suppose that at each $\vec{x}_a \in \mathcal{A}$ we obtain a noisy observation
\begin{align}
y_a = t(\vec{x}_a) + \varepsilon_a, \quad \varepsilon_a \sim \mathcal{N}(0, \sigma_a^2).  
\end{align}
Collecting all measurements gives the vector
\begin{align}
\label{measurement setup}
\mathbf{y}_{\mathcal{A}} = \mathbf{t}_{\mathcal{A}} + \boldsymbol{\varepsilon}_{\mathcal{A}} \sim \mathcal{N}\!\big(\mathbf{m}_{\mathcal{A}}, \, \mathbf{C}_{\mathcal{A}\mathcal{A}} + \boldsymbol{\Sigma}_{\mathcal{A}}\big),
\end{align}
where $\boldsymbol{\Sigma}_{\mathcal{A}}$ is the diagonal noise covariance matrix.  

\paragraph{Prediction at unsampled locations}  
For a location $\vec{x}^* \in \mathcal{S}\setminus\mathcal{A}$, the joint distribution of $\mathbf{y}_{\mathcal{A}}$ and $t(\vec{x}^*)$ is Gaussian, which leads to closed-form GP prediction equations \cite{Gp_book}:  
\begin{align}
&\mathbb{E}[t(\vec{x}^*) \mid \mathbf{y}_{\mathcal{A}}] = m(\vec{x}^*) 
+ \mathbf{c}_{\vec{x}^*\mathcal{A}} \big(\mathbf{C}_{\mathcal{A}\mathcal{A}} + \boldsymbol{\Sigma}_{\mathcal{A}}\big)^{-1} \big(\mathbf{y}_{\mathcal{A}} - \mathbf{m}_{\mathcal{A}}\big),&\label{eq:gp_mean}\\
&\mathrm{Var}[t(\vec{x}^*) \mid \mathbf{y}_{\mathcal{A}}] = c(\vec{x}^*,\vec{x}^*) 
- \mathbf{c}_{\vec{x}^*\mathcal{A}} \big(\mathbf{C}_{\mathcal{A}\mathcal{A}} + \boldsymbol{\Sigma}_{\mathcal{A}}\big)^{-1} \mathbf{c}_{\mathcal{A}\vec{x}^*},& \label{eq:gp_var}
\end{align}
where $\mathbf{c}_{\vec{x}^*\mathcal{A}}$ is the cross-covariance vector between $\vec{x}^*$ and the measurement set $\mathcal{A}$.  

\subsection{Design Objectives}

Having established the GP inference framework, we now outline the key design objectives that guide the proposed DT-assisted framework for measurement design and channel statistics.

\paragraph{Objective \#1- Estimation of geometry-enabled GP priors}  
Standard GP modeling often assumes a constant mean and imposes restrictive structures on the covariance function \cite{Gp_book}.  For example, a \emph{stationary} kernel depends only on the displacement $\vec{x} - \vec{x}'$, while an \emph{isotropic} kernel depends only on the Euclidean distance $\|\vec{x} - \vec{x}'\|_2$ .  
Common choices such as exponential kernels involve only a few tunable parameters.  
While convenient, these assumptions are frequently violated in wireless channels, where propagation is environment-dependent and correlations are highly non-stationary.  

In this work, we avoid such restrictive assumptions by leveraging an inaccurate DT of the environment, constructed using a ray-tracing model from open-source maps (e.g., OpenStreetMap).  
Without requiring precise material properties or geometry (i.e., using uncalibrated DTs), we estimate priors for both the mean $m(\vec{x})$ and covariance $c(\vec{x},\vec{x}')$ across candidate locations.  
This enables us to capture non-stationary and anisotropic spatial correlations induced by the environment, while retaining the GP framework for closed-form inference and uncertainty quantification.  
The first design question we address is therefore:  
\begin{quote}
\emph{How can we obtain realistic GP priors $(m(\cdot), c(\cdot,\cdot))$ for wireless channels without relying on overly simplistic kernel assumptions?}  
\end{quote}

\paragraph{Objective \#2 -- Measurement selection}
Equation~\eqref{eq:gp_var} illustrates an important property of GP-based inference, namely that the posterior variance does not depend on the actual measurement values.
This implies that GPs not only provide an expressive model for spatial prediction (interpolation) of channel statistics, but also offer a principled framework for selecting informative measurement locations through the choice of the measurement set $\mathcal{A}$.
Given the GP priors $(m,c)$ and the prediction space $\mathcal{S}$ obtained from the DT, the quality of predictions at unsampled locations $\mathcal{S}\setminus\mathcal{A}$ depends on the choice of measurement locations $\mathcal{A} \subseteq \mathcal{S}$.
The second design question is thus:
\begin{quote}
\emph{Which measurement locations $\mathcal{A}$ should be selected in order to optimize predictive accuracy across the remaining candidate positions?}
\end{quote}

\section{Digital Twin of the Wireless Channel}
\label{sec:DT Description}

\begin{figure}
    \centering
    \includegraphics[trim={8px 8px 8px 8px},clip,width=0.7\linewidth]{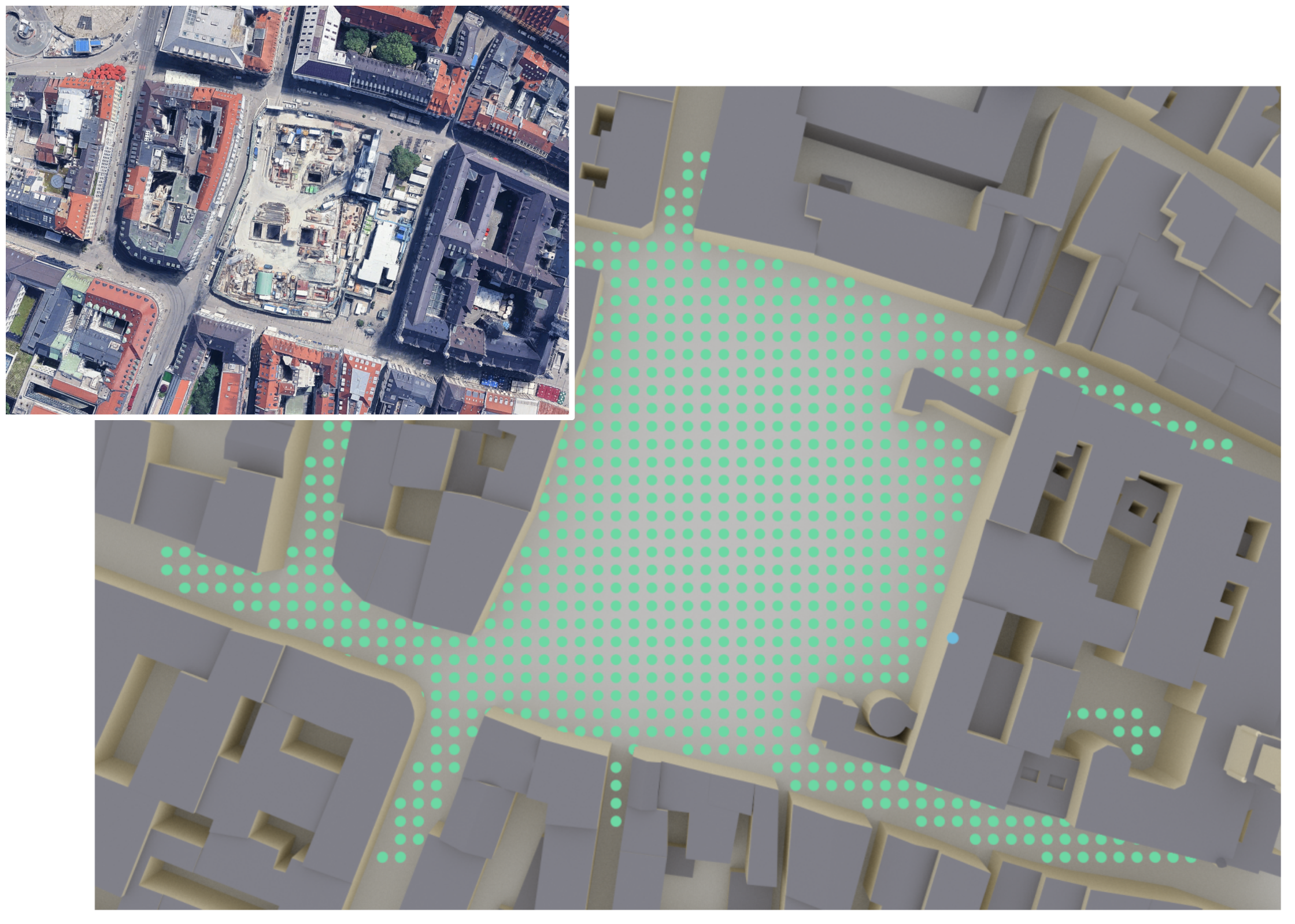}
    \caption{Illustrations of the Munich scene. Green dots indicate potential sensor location, blue dot indicates transmitter location. (satellite picture from google earth)}
    \label{fig:scenes}
\end{figure}

{A DT of the wireless channel consists of two main components: the physical scene and the ray tracing engine. Ray tracing is performed using Sionna version~0.19.2~\cite{Sionna_Intro}. The scene contains the geometric layout and electromagnetic material properties of the objects in the environment. The geometry is constructed from data obtained through the OpenStreetMap~\cite{OpenStreetMaps} API, which provides information about static elements such as buildings and roads. As a result, the generated scene does not include dynamic objects such as vehicles, or vegetation. An illustration of the scene is shown in Fig.~\ref{fig:scenes}. The accuracy of the provided geometry in terms of object size, shape, orientation, and position is uncertain, and some degree of inaccuracy must be accounted for. In conventional calibration procedures, the electromagnetic material properties are estimated through on-site measurements, which is a time-consuming process that must be repeated whenever the environment changes.}

As an alternative to the calibration process, this work instead treats the material parameters as unknown random variables. The uncertainty about the electromagnetic material parameters is modeled by randomly generating i.i.d. relative permittivities for all objects in the scene according to a uniform distribution $\epsilon_{\text{r}n} \sim \mathcal{U}\p{1.5, 30}$, where $\epsilon_{\text{r}} = 1.5$ corresponds to ceiling board and $\epsilon_{\text{r}} = 30$ corresponds to wet ground \cite{ituP2040}. The relative permeability is assumed to be $\mu_\text{r} = 1$ and the conductivity is assumed to be $\sigma_\text{c} = 0$ for all scene objects. The object positions are also randomized. The position of the $n$-th object is given as $\vec{v}_{n} = \veci{\upsilon}_n + \veci{\nu}_n$ where $\veci{\upsilon}_n \in \mathbb{R}^{3\times 1}$ is the position obtained through the OpenStreetMap API and the position errors $\veci{\nu}_n$ {are randomly drawn to model the geometry uncertainty}. In the following, $\veci{\beta}$ denotes the material parameters and object positions.

The scene is populated with an AP placed at a position denoted $\vec{x}_\text{AP}$ at a height of 27 meters. A square grid of candidate sensors is placed in the scene. The sensor grid has a spacing of 5 meters and is placed at a height of 1.5 meters. After the pruning process, the scene has $M=1002$ candidate sensor locations, $\mathcal{S} = \{\vec{x}_m\}^{M}_{m=1}$, where $\vec{x}_m \in \mathbb{R}^{3\times 1}$. The AP and all receivers are equipped with a single vertical half-wavelength dipole.

 {Ray traced channel estimates for each link between the transmitter and a sensor location are obtained by generating permittivities $\epsilon_{\text{r}k}$, applying position shifts $\veci{\nu}_k$, and running the ray tracing algorithm.}
 In this work,  the ray tracer is configured for a center frequency at $6 \text{ GHz}$, $8 \text{ GHz}$ double-sided bandwidth, subcarrier spacing $1 \text{ MHz}$, $N=8001$ subcarriers, and $10^6$ rays per source. {The DT thus constitutes a function that maps a vector  $\veci{\beta}$ containing material properties and other scene information (i.e., object positions) to a channel power matrix} $\mat{P}\p{\veci{\beta}} = \begin{bmatrix}
\vec{p}_1\p{\veci{\beta}} &  \cdots & \vec{p}_M\p{\veci{\beta}}
\end{bmatrix}$, where $\vec{p}_m\in \mathbb{R}^{N\times1}$ denotes the power of the channel between the transmitter and the $m$'th sensor location.

\section{Proposed Digital Twin-Enabled Probing and Channel Statistics Prediction Scheme}
\label{sec: Proposed DT-Enabled Probing and Channel Statistics Prediction Scheme}

This section presents the proposed DT-enabled probing and channel statistics prediction framework. 
Building upon the GP formulation introduced in Section~\ref{sec:Problem Statement}, the goal is to predict location-dependent channel statistics with minimal channel measurements by exploiting spatial correlations informed by the uncalibrated DT of the environment introduced in Section~\ref{sec:DT Description}.

\subsection{Channel Statistics and Processing}
\label{subsection:processing}
We begin by defining a suitably transformed representation of the channel power quantile, whose logarithmic form exhibits approximately Gaussian statistics, thereby lending itself naturally to GP-based inference. 
Let $\mathcal{\hat{D}}=\{(\vec{x}_m,\vec{p}_m)\}_{m=1}^{M}$ denote the measurement dataset associated with candidate sensor locations, where each $\vec{p}_m = (p_{m,1},\ldots,p_{m,N})$ contains $N$ realizations of the small-scale fading power observed at location $\vec{x}_m$. For each location, the $\epsilon$-quantile of the channel power is estimated empirically as $\widehat{p}_{\epsilon,m} = p_{m,(r)}, r = \lfloor N \epsilon \rfloor,$
where $p_{m,(r)}$ denotes the $r$-th order statistic of $\vec{p}_m$. 
Applying a logarithmic transformation yields
\begin{align}
t(\vec{x}_m) \triangleq \widehat{q}_{\epsilon,m} = \log(\widehat{p}_{\epsilon,m}), \label{eq:quantile_compute}
\end{align}
which provides a consistent and asymptotically Gaussian estimate of the log-quantile channel power $t(\vec{x}_m)$~\cite{GP_radio_map_3}. 
This transformation, consistent with the general formulation $t(\vec{x}) = g(p_\epsilon(\vec{x}))$ introduced in \eqref{transform_stat}, allows the resulting process $t(\vec{x})$ to be effectively modeled as a GP \cite{GP_radio_map_3,GP_radio_map_2}. 
The transformed dataset is thus expressed as $\mathcal{D} = \{ (\vec{x}_m, \widehat{q}_{\epsilon,m}) \}_{m=1}^{M},$
and serves as the input to the subsequent DT-aided GP inference and sensor selection stages.

\subsection{Estimation of Gaussian Process Priors Using the Digital Twin}
\label{sec: Estimation of GP Priors}

The DT provides a geometry-aware prior over both the mean and covariance functions of the GP, enabling the fusion of simulated and channel measurements for reliable spatial interpolation and uncertainty quantification. 
Instead of relying on restrictive kernel assumptions, the proposed approach leverages the uncalibrated DT described in Section~\ref{sec:DT Description} to derive spatially varying priors that reflect the underlying propagation geometry. 
This integration allows the GP model to capture complex, non-stationary, and anisotropic correlations induced by the environment, while maintaining the tractability and interpretability of the GP framework for closed-form inference and uncertainty estimation illustrated in \eqref{eq:gp_mean} and \eqref{eq:gp_var}.

Following the hierarchical modeling framework of~\cite{kennedy2001bayesian}, uncertainty introduced by the DT due to imperfect object positions and unknown material properties is treated as a higher-level source of randomness governing the GP prior. {We construct DT-informed GP priors by directly matching the first- and second-order moments
of the spatial field induced by DT randomization.
Specifically, the DT is used to generate realizations of the transformed channel statistic across the prediction space by randomizing object positions and material properties in the ray tracer.
These realizations are treated as samples of the latent spatial process, from which the mean and covariance are defined as
\begin{subequations}
    \begin{align}
    \vec{m}_{\mathcal{S}} &= \mathbb{E}_{\boldsymbol{\beta}}\!\big[\vec{q}(\boldsymbol{\beta})\big], \label{mean est}\\
    \mathbf{C}_{\mathcal{S}\mathcal{S}} &=
    \mathbb{E}_{\boldsymbol{\beta}}\!\Big[
    \big(\vec{q}(\boldsymbol{\beta})-\vec{m}_{\mathcal{S}}\big)
    \big(\vec{q}(\boldsymbol{\beta})-\vec{m}_{\mathcal{S}}\big)^{\!H}
    \Big], \label{covariance est}
    \end{align}
\end{subequations}
where the expectations are approximated via Monte Carlo sampling of the DT with
$\boldsymbol{\beta}$ drawn as described in Section~\ref{sec:DT Description}.
Here, $\vec{q}(\boldsymbol{\beta}) =
\big[\widehat{q}_{\epsilon,1}, \widehat{q}_{\epsilon,2}, \ldots, \widehat{q}_{\epsilon,M}\big]^{\!T}$
denotes a realization of the transformed channel statistic over the prediction space $\mathcal{S}$,
with $\widehat{q}_{\epsilon,m}$ defined in \eqref{eq:quantile_compute}.
This moment-matched construction yields an effective GP prior that can capture geometry-induced, non-stationary correlations without assuming kernel smoothness or stationarity, while preserving closed-form GP inference.} 

\subsection{Optimized Selection of Channel Probing}
\label{sec:MI_selection}

{Given GP priors $(m(\cdot), c(\cdot,\cdot))$ and the prediction space $\mathcal{S}$, the quality of predictions at unsampled locations $\mathcal{S}\setminus\mathcal{A}$ depends on the set of measurement locations $\mathcal{A} \subseteq \mathcal{S}$.  
As shown in \eqref{eq:gp_var}, the posterior variance depends only on the choice of $\mathcal{A}$ and not on the actual observed values, making it a natural criterion for optimizing measurement design.}  

{We leverage the DT as a geometric prior to guide measurement placement.  
Even if the DT is uncalibrated or exhibits mismatched material or positional parameters, its geometry induces a covariance structure that can be exploited for informative sampling. Using the DT-based covariance estimate of \eqref{covariance est} as a prior, our goal is to select a subset $\mathcal{A} \subseteq \mathcal{S}$ whose measurements yield the greatest reduction in uncertainty across the remaining unsampled points $\mathcal{S}\setminus\mathcal{A}$.}  

\paragraph{Mutual information objective}
{Let $t_{\mathcal{S}}$ denote the channel statistic of interest $t(\vec{x})$ over the prediction space of $\vec{x} \in \mathcal{S}$.}  
We seek to maximize the mutual information (MI) between the sampled and unsampled points:
\begin{align}
\label{eq:MI_obj_revised}
 \mathcal{A}^* = \arg\max_{\mathcal{A} \subseteq \mathcal{S}, |\mathcal{A}| = k} I(\mathbf{y}_{\mathcal{A}}; t_{\mathcal{S}\setminus\mathcal{A}})
, 
\end{align}
where the mutual information $I(\mathbf{y}_{\mathcal{A}}; t_{\mathcal{S}\setminus\mathcal{A}})
= H(t_{\mathcal{S}\setminus\mathcal{A}})
- H(t_{\mathcal{S}\setminus\mathcal{A}} \mid \mathbf{y}_{\mathcal{A}}),
$ quantifies how much knowing the measurements at 
$\mathcal{A}$ reduces uncertainty about the unsensed region $\mathcal{S} \setminus \mathcal{A}$. In essence, the goal is to select sensor placements that are most informative about unsensed locations.

\paragraph{Greedy approximation algorithm}
Since solving \eqref{eq:MI_obj_revised} exactly is combinatorial and NP-hard \cite{Krause_2008}, we adopt a simple but effective greedy algorithm.  
Starting with $\mathcal{A}_0 = \emptyset$, at each iteration $j$ and with candidate locations $\tilde{\vec{x}} \in \mathcal{S}\setminus\mathcal{A}_{j-1}$,   the next measurement location $\vec{x}^*$ is chosen to maximize the marginal information gain:
    \begin{align}
 &\vec{x}^* = \arg\max_{\tilde{\vec{x}} \in \mathcal{S}\setminus\mathcal{A}_{j-1}} 
\Delta(\tilde{\vec{x}} \mid \mathcal{A}_{j-1}), 
\quad 
\mathcal{A}_j = \mathcal{A}_{j-1} \cup \{\vec{x}^*\}, \nonumber
\\ 
 &\Delta(\tilde{\vec{x}} \mid \mathcal{A}) 
 = H(t_{\tilde{\vec{x}}} \mid \mathbf{y}_{\mathcal{A}})
 - H(t_{\tilde{\vec{x}}} \mid t_{\bar{\mathcal{A}}}),
,
\end{align} \label{eq:selection}
and $\bar{\mathcal{A}} = \mathcal{S}\setminus(\mathcal{A}\cup\{\tilde{\vec{x}}\})$. Under GP setup, this expression of the mutual information (MI) simplifies to
\begin{align}
\Delta (\tilde{\vec{x}} \mid \mathcal{A})
= \frac{1}{2}
\log\!\left(
\frac{
\mat{C}_{\{\tilde{\vec{x}}\}\{\tilde{\vec{x}}\}}
- \mat{C}_{\{\tilde{\vec{x}}\}\mathcal{A}}
(\mat{C}_{\mathcal{A}\mathcal{A}} + \boldsymbol{\Sigma}_{\mathcal{A}})^{-1}
\mat{C}_{\mathcal{A}\{\tilde{\vec{x}}\}}
}{
\mat{C}_{\{\tilde{\vec{x}}\}\{\tilde{\vec{x}}\}}
- \mat{C}_{\{\tilde{\vec{x}}\}\bar{\mathcal{A}}}
\mat{C}^{-1}_{\bar{\mathcal{A}}\bar{\mathcal{A}}}
\mat{C}_{\bar{\mathcal{A}}\{\tilde{\vec{x}}\}}
}
\right),
\end{align}
where $\mat{C}_{\mathcal{I}\mathcal{J}}\, | \,\mathcal{I}, \mathcal{J} \in \{\tilde{\vec{x}}, \mathcal{A},\bar{\mathcal{A}} \}$ is a sampling of the respective rows and columns of $\mat{C}_{\mathcal{S}\mathcal{S}}$, defined in \eqref{covariance est}. This selection algorithm prioritizes measurement points that are both uncertain given previous observations and representative of unmeasured locations, ensuring  coverage with minimal redundancy \cite{Krause_2008}.  

\paragraph{Submodularity and performance guarantee}
{As established in \cite{Krause_2008}, the MI criterion for GPs is monotone and submodular, meaning that additional measurements yield diminishing marginal gains:
\begin{align}
\Delta(\tilde{\vec{x}} \mid \mathcal{A}) \ge \Delta(\tilde{\vec{x}} \mid \mathcal{B}), 
\quad \forall\, \mathcal{A} \subseteq \mathcal{B}, \; \tilde{\vec{x}} \notin \mathcal{B}.
\end{align}
Consequently, the greedy algorithm provides a $(1 - 1/e)$-approximation to the optimal solution.  
Subsequently, this ensures that, even when the DT prior is imperfect and a simple greedy algorithm is employed, the proposed selection procedure remains efficient and theoretically grounded, leveraging the DT’s geometric structure to guide informative measurements.
}

\emph{Proposed Scheme:}
Given the GP prior information, i.e., $\vec{m}_\mathcal{S}$ and $\mat{C}_{\mathcal{S}\mathcal{S}}$ estimated from the DT as described in Section~\ref{sec: Estimation of GP Priors}, and the  selection framework of Section~\ref{sec:MI_selection} providing the measurement points $\mathcal{A} \subseteq \mathcal{S}$, the log-power quantiles are probed (i.e., measured) at the selected points $\mathcal{A}$, forming the observation vector $\vec{y}_\mathcal{A}$. Given the observation vector $\vec{y}_\mathcal{A}$ together with the prior mean vector $\vec{m}_\mathcal{S}$ and covariance matrix $\mat{C}_{\mathcal{S}\mathcal{S}}$, the closed-form GP prediction equations in~\eqref{eq:gp_mean} and~\eqref{eq:gp_var} are applied to compute the posterior mean $\widehat{\vec{m}}_\mathcal{S}$ and variance $\widehat{\vec{c}}_\mathcal{S}$ for all points in $\mathcal{S}$. Here, the sub-vectors and sub-matrices $\vec{m}_\mathcal{A}$ and $\mat{C}_{\mathcal{A}\mathcal{A}}$ are sampled from $\vec{m}_\mathcal{S}$ and $\mat{C}_{\mathcal{S}\mathcal{S}}$, respectively.
{While the repeated ray tracing of the scene required to build the priors incurres computational overhead, these computations can be run offline and results stored for online use. Additionally, after selection of channel probing locations, the matrix inverse $\p{\mat{C}_{\mathcal{A}\mathcal{A}}+ \vec{\Sigma}_{\mathcal{A}}}^{-1}$ can be precomputed. During online operation, the channel prediction Eqs. \eqref{eq:gp_mean} and \eqref{eq:gp_var} then only involves the computation of matrix-vector products.  }

\section{Numerical Results}\label{sec:simulation}
In this section, we demonstrate the efficacy of the proposed DT-enabled probing and channel statistics prediction framework presented in Section~\ref{sec: Proposed DT-Enabled Probing and Channel Statistics Prediction Scheme}. For the scene shown in Fig.~\ref{fig:scenes} and the setup described in Section~\ref{sec:DT Description}, the performance of the proposed scheme is evaluated at points $\vec{x}^* \in \mathcal{S}$ {using two} metrics:

\begin{enumerate}[labelwidth=*, labelindent=0pt, leftmargin=*]
    \item \textbf{Mean absolute error of prediction}, defined as
    \begin{align}
    \text{mean}\p{\abs{\text{error}}} =M^{-1} \mat{1}_{M\times 1}\T \abs{\widehat{\vec{m}}_\mathcal{S} - \vec{q}(\hat{\veci{\beta}})}.
    \end{align}
    where $\hat{\veci{\beta}}$ is a realization drawn according to Sec. \ref{sec:DT Description}, serving as the target for estimation.
    
    \item \textbf{Normalized rate under meta-probability constraint:} the prediction scheme is applied to satisfy the meta-probability constraint in~\eqref{ref: meta prob 1} with confidence parameter $\delta = 0.05$, while optimizing the normalized rate $= \frac{R(\vec{x}^*)}{R_{\sf{ideal}}(\vec{x}^*)},$
    where, for predictive mean $\mu(\vec{x}^*) \overset{\Delta}{=} \mathbb{E}[t(\vec{x}^*) \mid \mathbf{y}_{\mathcal{A}}]$ given by~\eqref{eq:gp_mean} and predictive variance $\sigma^2(\vec{x}^*) \overset{\Delta}{=} \mathrm{Var}[t(\vec{x}^*) \mid \mathbf{y}_{\mathcal{A}}]$ given by~\eqref{eq:gp_var}, the achievable rate is selected as~\cite{GP_radio_map_2, GP_radio_map_5}
    \begin{align}
        R(\vec{x}^*) = \log_2 \Big( 1 + \e{\mu(\vec{x}^*) + \sqrt{2}\, \sigma(\vec{x}^*) \erf^{-1}(2 \delta - 1)} \Big). \label{eqn: rate selection}
    \end{align}  
    Here, $R_{\sf{ideal}}(\vec{x}^*)$ denotes the outage capacity computed using the ground truth fading distribution, i.e., assuming ideal CSI knowledge.
\end{enumerate}

We evaluate the performance of the proposed scheme, described in Section~\ref{sec: Proposed DT-Enabled Probing and Channel Statistics Prediction Scheme}, {with GP prior built using 50 realizations of $\veci{\beta}$. For each of the 50 realizations, the position errors are i.i.d. and drawn from a uniform distribution, $\veci{\nu}_n \sim \mathcal{U}\p{\begin{bmatrix} -2 & -2& 0\end{bmatrix}\T, \begin{bmatrix} 2 & 2 & 0\end{bmatrix}\T }$.  The proposed scheme is evaluated against} the following benchmarks:

\begin{enumerate}[labelwidth=*, labelindent=0pt, leftmargin=*]
    \item \emph{Uninformed GP baseline:}   $\mathcal{A}$ is selected at random, uniformly, from $\mathcal{S}$. The mean function $\vec{m}_\mathcal{S} = \mat{0}_{M\times 1}$ and the covariance matrix $\mat{C}_{\mathcal{S}\mathcal{S}}$ is given by the Matern kernel. The hyper-parameters are obtained by maximizing the likelihood of the observed data with labels as  $(\vec{y}_\mathcal{A} - \vec{m}_\mathcal{A})$ and features as $\mathcal{A}$, following the procedure in \cite[Ch. 5]{Gp_book}.

    \item \emph{Stationary DT–GP baseline:} In this scheme, only a single realization of $\veci{\beta}$ is utilized. The mean function is obtained by linear regression. In wireless systems, the mean function of the channel power is often approximated by an exponential function \cite[Sec. 2.5]{goldsmith_wireless_2005}. In logarithmic scale, the exponential function is {$\tilde{m}\p{\vec{x}} = \zeta + \alpha \log\p{d\p{\vec{x}}},$}
where $d\p{\vec{x}} = \norm{\vec{x} - \vec{x}_\text{Tx}}_2$ is the euclidean distance to the transmitter and the constants $\alpha$, $\zeta$ are obtained through linear regression
; $\begin{bmatrix}
\zeta & \alpha
\end{bmatrix}\T = \p{\mat{A}\H \mat{A}}^{-1} \mat{A}\H \vec{q}$$(\veci{\beta}),$
where $\mat{A} = \begin{bmatrix}
1 & \cdots & 1 \\
\log\p{d\p{\vec{x}_1}} & \cdots & \log\p{d\p{\vec{x}_M}}
\end{bmatrix}\T$. The mean vector can then be estimated as $\vec{m}_\mathcal{S} = \tilde{m}\p{\mathcal{S}}$. For the selection algorithm, an estimate of the covariance matrix $\mat{C}_{\mathcal{S}\mathcal{S}}$ is needed. This estimate is obtained by the Matern kernel fitted to the labels $(\vec{q}(\veci{\beta}) - \vec{m}_\mathcal{S})$ and features $\mathcal{S}$.
For the GP prediction scheme, a new realization of the covariance matrix $\mat{C}_{\mathcal{S}\mathcal{S}}$ is given by fitting the Matern kernel to the observed labels $(\vec{y}_\mathcal{A} - \vec{m}_\mathcal{A})$ and features $\mathcal{A}$.
\end{enumerate}

\begin{figure}[]

        \centering    \includegraphics[trim={0px 30px 0px 0px},clip,width=1.05\linewidth]{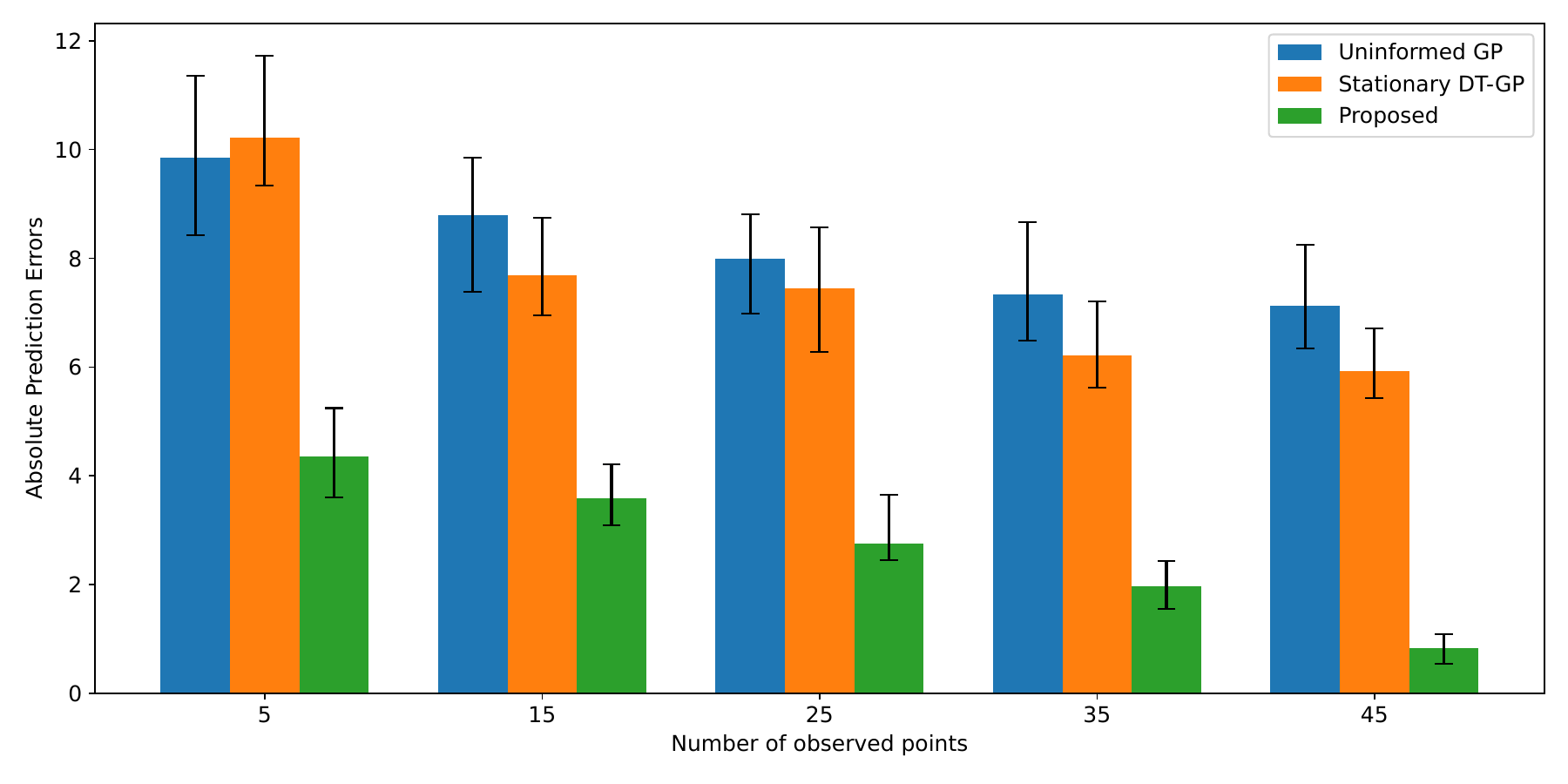}
    \caption{Mean absolute error of the GP as a function of observed points.}
    \label{fig:mean_error}
\end{figure}

{Figure~\ref{fig:mean_error} shows the mean absolute prediction error of the estimated channel statistic as a function of the number of observed measurement locations. The results report the median error and the $75\%$ confidence interval over 50 independent realizations for three schemes: an uninformed GP baseline, a DT-assisted GP with stationary kernel fitting, and the proposed geometry-aware GP with DT-estimated priors. As the number of observed points increases, all methods initially benefit from additional measurements; however, the proposed scheme consistently achieves substantially lower prediction error across all measurement budgets. In particular, while the benchmark methods exhibit a residual error floor even with many observations, the proposed approach continues to reduce error as more measurements are collected. This behavior indicates that geometry-aware priors derived from uncalibrated DTs enable the GP to capture environment-induced spatial correlations that cannot be adequately learned through stationary kernel fitting alone, thereby improving prediction accuracy even in the high-measurement regime.}

\begin{figure}[t]
    \centering
    \includegraphics[trim={0px 30px 0px 0px},clip,width=0.9\linewidth]{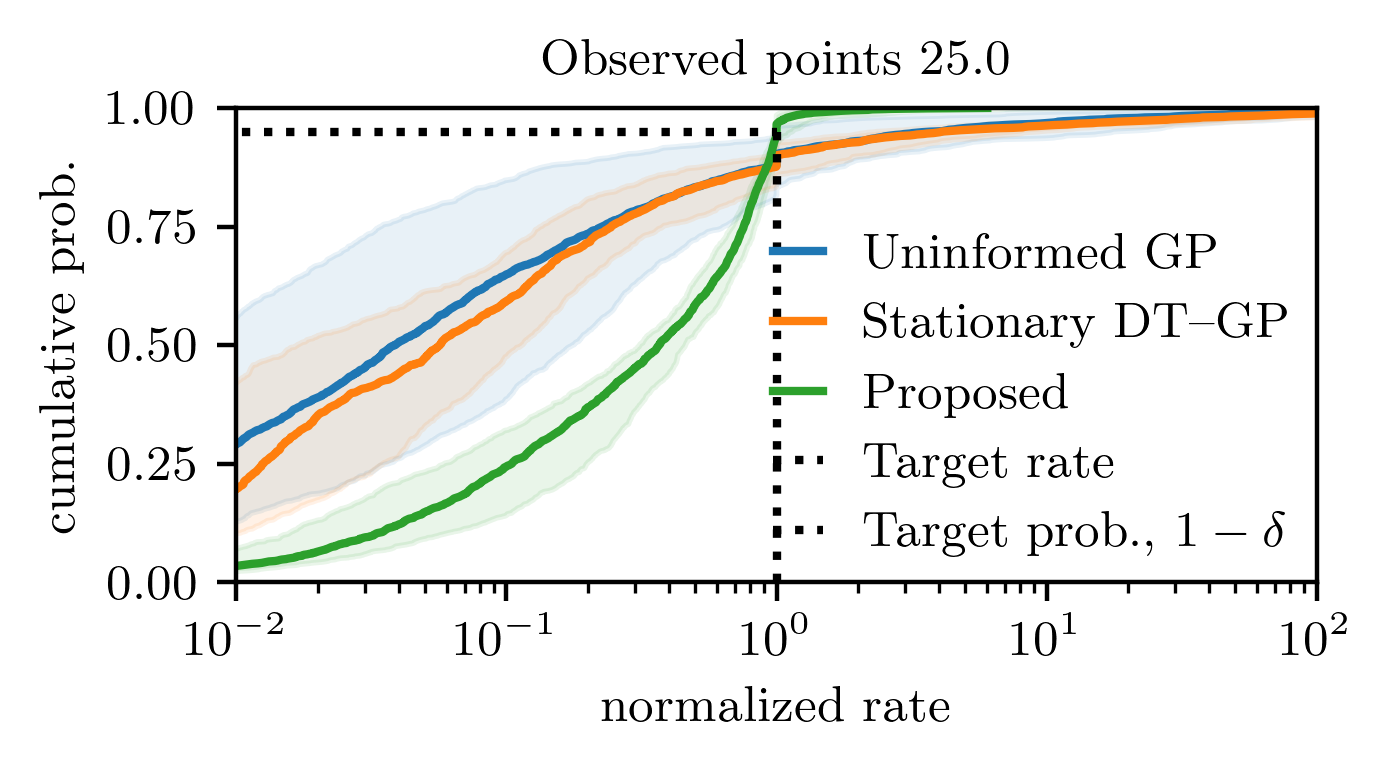}\\
    \includegraphics[trim={0px 30px 0px 0px},clip,width=0.9\linewidth]{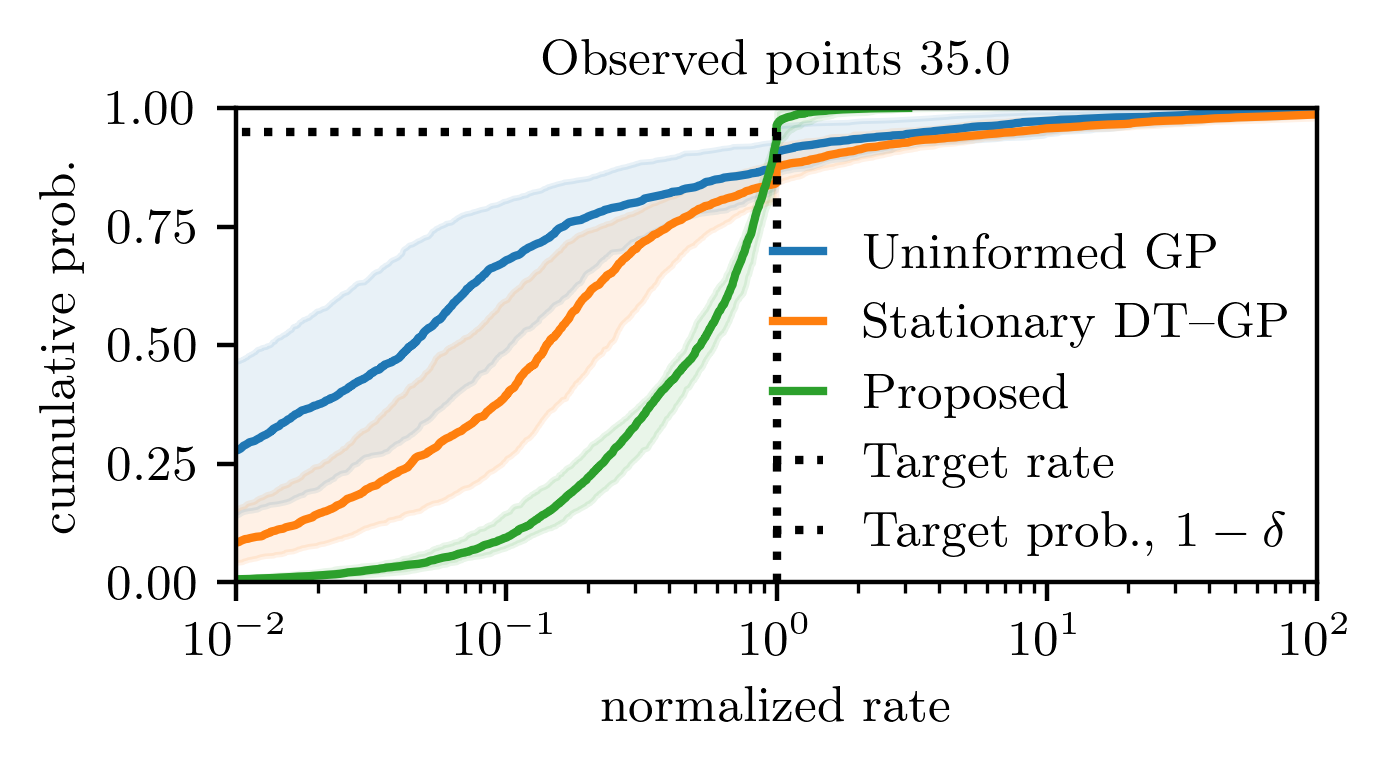}
      \includegraphics[trim={0px 0px 0px 0px},clip,width=0.9\linewidth]{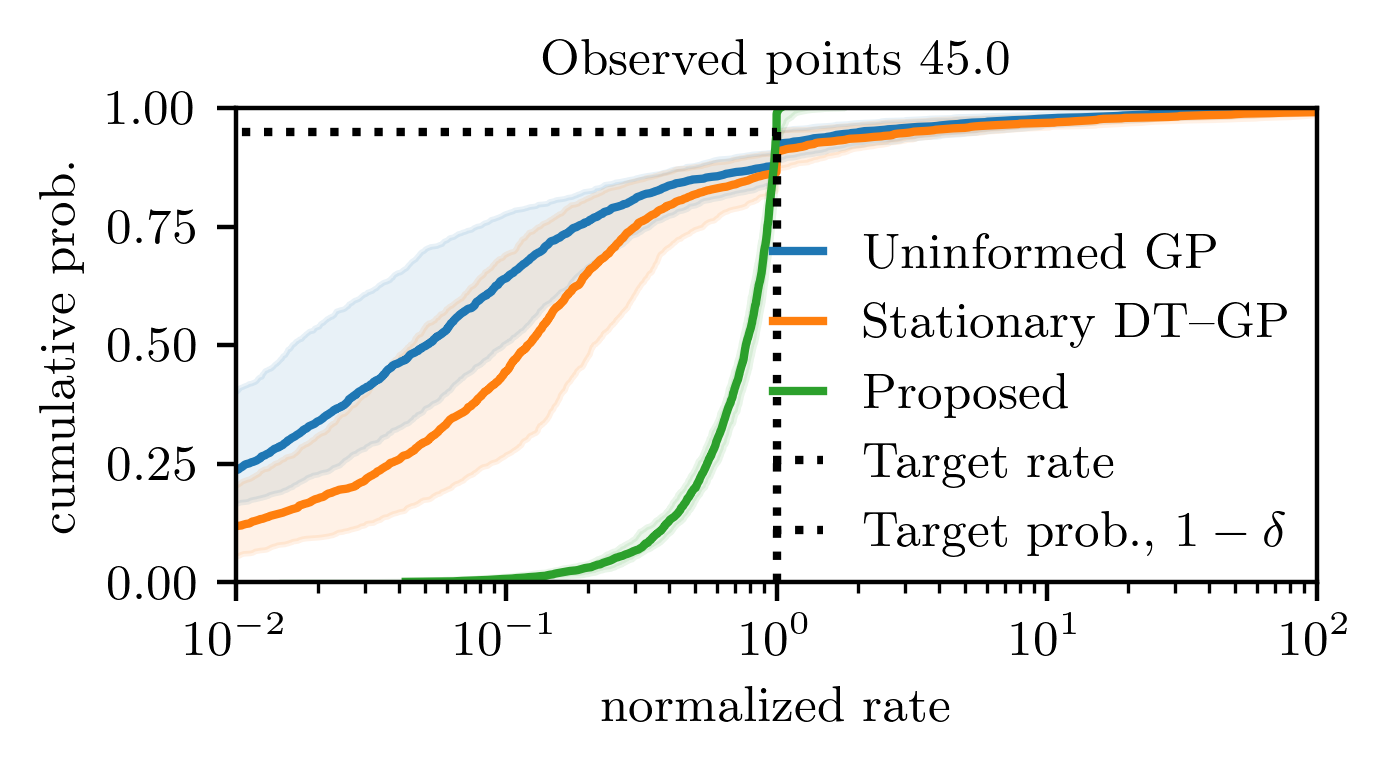}
    \caption{CDF of the normalized rate for 25, 35, and 45 observed points.}
    \label{fig:rate_munich}
\end{figure}

{Figure~\ref{fig:rate_munich} illustrates the CDF of the normalized rate under the meta-probability constraint for 25, 35, and 45 observed measurement locations. The black dotted lines indicate the target operating point defined by the required confidence level and the corresponding normalized rate. Across different measurement budgets, only the proposed scheme, with rate selection according to \eqref{eqn: rate selection}, satisfies the target meta-probability, as indicated by the intersection of the two dotted lines. Notably, this is achieved while maintaining a higher normalized rate than the benchmark methods, as evidenced by the proposed scheme lying to the right of the benchmark curves. As the number of observed points increases, this joint satisfaction of reliability and rate optimality leads to a progressively sharper transition in the CDF, approaching the step-function behavior expected when the underlying channel statistics are accurately captured. In contrast, the benchmark methods exhibit a more gradual CDF transition and do not align with the target operating point.}

\section{Conclusions}\label{sec:conclusion}

{This paper presented a hybrid framework for spatial prediction of wireless channel statistics that integrates uncalibrated digital twins with Gaussian process regression to achieve geometry-aware and data-efficient inference. By estimating geometry-induced priors for both the mean and covariance directly from imperfect digital twins, without requiring material calibration, the proposed approach embeds environmental structure into the inference process rather than relying solely on kernel-based fitting from measurements. Numerical evaluations demonstrate that this enables the Gaussian process model to capture complex, environment-induced spatial correlations that are poorly represented by geometry-unaware or stationary kernel-based benchmarks, resulting in more accurate and reliable prediction of channel statistics with significantly fewer measurements. These gains translate into improved statistical guarantees for URLLC rate selection, highlighting how uncalibrated digital twins can serve as a practical source of structural prior information for data-efficient and site-specific channel statistics prediction in URLLC scenarios.}

\bibliographystyle{IEEEtran}
\bibliography{references.bib}
\end{document}